\documentclass[sigconf]{acmart}
\acmArticle{1}

\setcopyright{none}
\renewcommand\footnotetextcopyrightpermission[1]{}
\AtBeginDocument{%
  }

\setcopyright{acmcopyright}
\acmConference[The ACM SIGGRAPH / Eurographics Symposium on Computer Animation 2023]{August 04--06, 2023}{Los Angeles, California}

\begin{document}

\title{ACT2G: Attention-based Contrastive Learning \\ 
      for Text-to-Gesture Generation}


% \author{
% Submission 9070
% }

\author{Hitoshi Teshima}
\orcid{0000-0002-6431-4514}
\authornote{teshima.hitoshi.058@s.kyushu-u.ac.jp}
\affiliation{%
  \institution{Kyushu University}
  \city{Fukuoka}
  \country{Japan}
}

\author{Naoki Wake}
\orcid{0000-0001-8278-2373}
\affiliation{%
  \institution{Microsoft}
  \city{Redmond}
  \state{Washington}
  \country{USA}
}

\author{Diego Thomas}
\orcid{0000-0002-8525-7133}
\affiliation{%
  \institution{Kyushu University}
  \city{Fukuoka}
  \country{Japan}
}

\author{Yuta Nakashima}
\orcid{0000-0001-8000-3567}
\affiliation{%
  \institution{Osaka University}
  \city{Osaka}
  \country{Japan}
}

\author{Hiroshi Kawasaki}
\orcid{0000-0001-5825-6066}
\affiliation{%
  \institution{Kyushu University}
  \city{Fukuoka}
  \country{Japan}
}

\author{Katsushi Ikeuchi}
\orcid{0000-0001-9758-9357}
\affiliation{%
  \institution{Microsoft}
  \city{Redmond}
  \state{Washington}
  \country{USA}
}


\begin{abstract}
Recent increase of remote-work, online meeting and tele-operation task makes 
people find that gesture for avatars and communication robots is more 
important than we have thought. It is one of the key factors to achieve smooth and natural 
communication between humans and AI systems and has been intensively researched.
Current gesture generation methods are mostly based on deep neural network using text, audio and other information as the input, 
however, they generate gestures mainly based on audio, which is called a beat gesture.
Although the ratio of the beat gesture is more than 70\% of actual human gestures, content based gestures sometimes play an important role to make avatars more realistic and human-like.
In this paper, we propose a attention-based contrastive learning for text-to-gesture (ACT2G), where 
 generated gestures represent content of the text by estimating attention weight for each word from 
the input text.
In the method, since text and gesture features calculated by the 
attention weight are mapped to the same latent space by contrastive 
learning, once text is given as input, the network outputs a 
feature vector which can be used to generate gestures related to the 
content. User study confirmed that the gestures generated by ACT2G were 
better than existing methods. In addition, it was demonstrated that wide variation of gestures were generated from the same text by changing attention weights by creators.

\ifx
Recently, importance of gesture for avatars and communication robots 
is noticed and intensively researched.
Current gesture generation methods are mostly neural network based and use text, audio, or other modalities as input and output a sequence of postures. However, most existing methods generate monotone gestures synchronized with audio information, called beat gesture, but ignore content information, called imagistic gesture, which makes generated gestures unrealistic. 
To generate gestures based on contents,
we propose a new framework named Attention-based Contrastive Learning for Text-to-Gesture (ACT2G), where gestures are generated by estimating attention weight for each word from the input text, while selected words are likely to express the gesture. For gesture generation, text and gesture features calculated by the attention weight are mapped to the same latent space by contrastive learning,
where gestures are categorized into the limited number of groups by clustering in advance.
Since feature vectors are mapped into the same space, once text is given as input, the network outputs a feature vector which can be used to generate gestures related to the content. 
User study confirmed that the gestures generated by ACT2G were better than existing methods in terms of human-likeliness and appropriateness. It was also demonstrated that diverse gestures were generated by providing arbitrary attention weights by creators.
\fi
\end{abstract}



\begin{teaserfigure}
    % \vspace{3.5cm}
     \includegraphics[width=\linewidth]{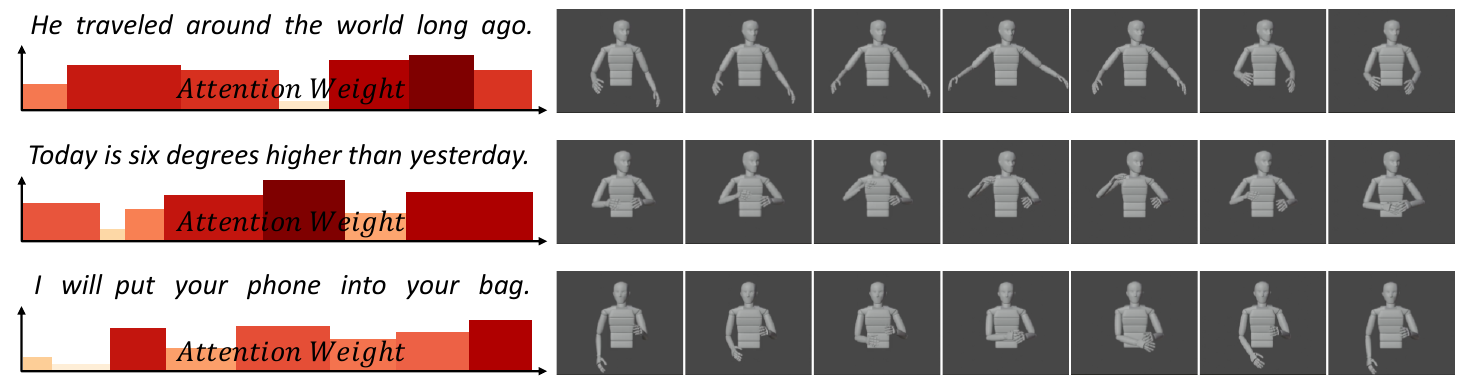}
     \centering
      \caption{ACT2G takes text as input and outputs realistic gestures. Text is encoded based on the Attention Weight, which represents the likelihood that the gesture will appear, and gesture is generated from the text feature.}
    \label{fig:overview}
    \vspace{0.5cm}
\end{teaserfigure}


\maketitle

\section{Introduction}

In recent years, communication in virtual space has become more active and avatars are increasingly used. In addition tele-operation robots and communication robots become popular and widely developed. 
Past psychological research has shown that gestures play an important role in conveying information~\cite{MCN94, RLB10}, but gesturing avatars and robots is one big challenge. Since manual design of gesture is time-consuming, %as a consequence 
gesture generation methods have been actively studied for long time. However, %in recent years, however,
%
%gesture generation methods based on data-driven approach have been actively studied in recent years~\cite{YKJ19, A2G21, S2G19, TDN21, qian2021speech}. Many existing methods use audio, text, or both as input, but 
it has been extremely difficult to properly reflect the meaning of the content of speech by previous rule based method because the relationship between the semantic information in the speech and the gesture has not been considered.
%yet in previous approaches.

To solve the problem, learning based approaches have been 
proposed~\cite{YKJ19, A2G21, S2G19, TDN21, qian2021speech}, where gestures are 
generated from audio or text information trained by using real gesture databases, expecting semantic information being implicitly considered. 
However, most existing methods generate gestures mainly based on audio, which is called a beat gesture, because the ratio of the beat gesture is more than 70\% of actual human gestures~\cite{MCN94}.
It should be noticed that text/content based gesture sometimes plays an important role to make gestures more realistic and more human-like.
%However, variation of gesture is 
%extremely large because they are mixed with a) beat gestures, which do not carry any 
%semantic information, but only correlated with audio, b) representational gestures, which 
%express semantic information correlated with text, and c) random gestures. 

\begin{figure}[t]
    \centering
    \includegraphics[width=\linewidth]{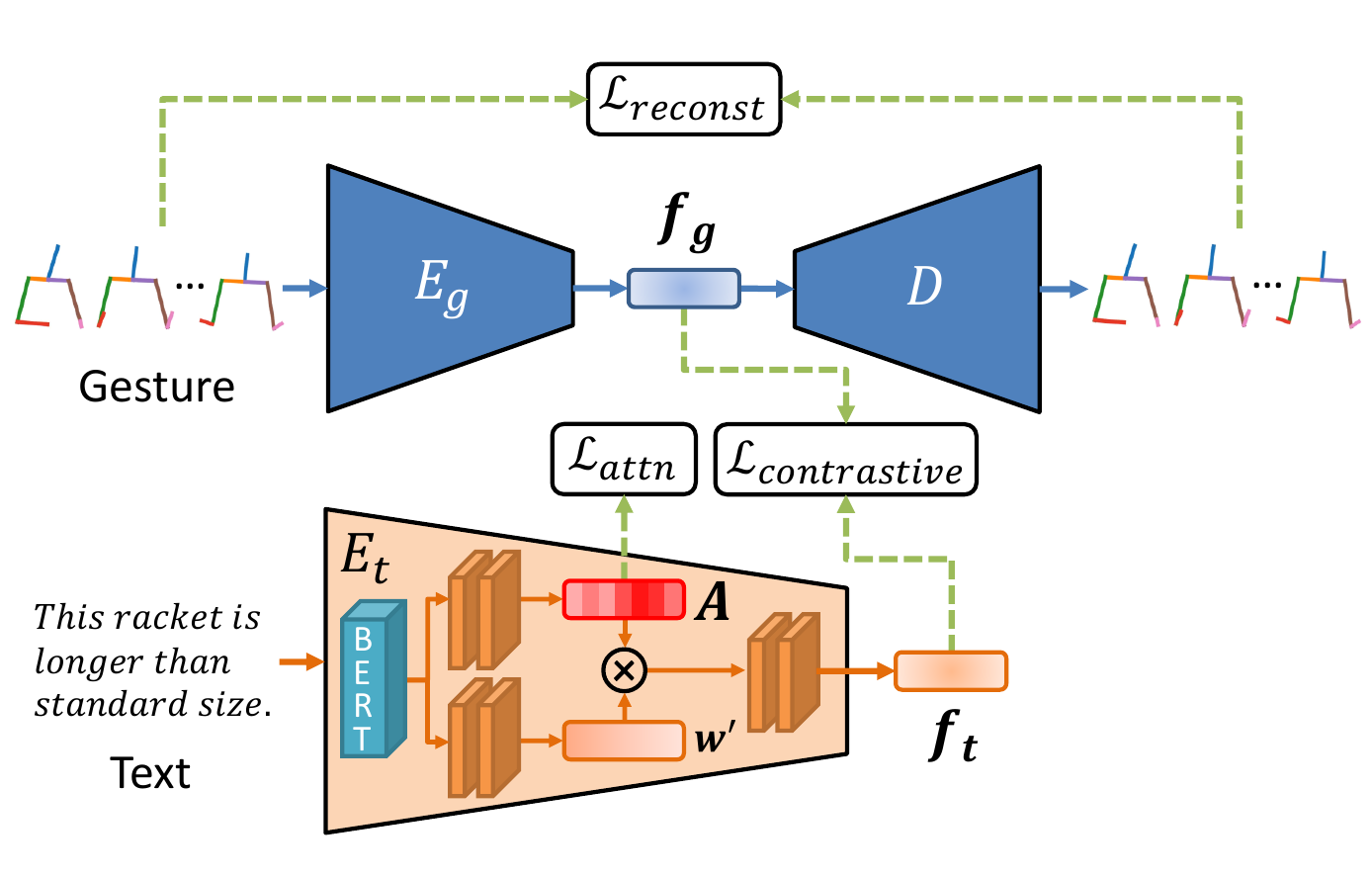}
    \caption{Pipeline for training. Encoded text feature $\textbf f_t$ and gesture feature $\textbf f_g$ are mapped into a multimodal space by contrastive learning.}
    \label{fig:network}
\end{figure}

In the paper, 
we propose Attention-based Contrastive Learning for Text-to-Gesture (ACT2G), 
which is a pipeline to generate gestures only from text and explicitly represents 
the semantic information as shown in Fig. \ref{fig:network}.
In our technique, to generate the large variation of gestures from arbitrary text,  VAE is applied to 
encode the sequence of gestures into small dimension, and then, texts encoded by 
a Transformer network  are mapped to the same latent space by
%clustering technique is applied to make a limited number of representative gestures.
%Then, to explicitly correlate text to gesture, 
contrastive learning, %is applied
%where textual features encoded by a Transformer network are mapped to the same latent space of gestures encoded by VAE. This 
%to mapp them to the same latent space by VAE. This 
%contrastive learning is effective to correlate 
%semantic information to gestures.
by which semantic information is effectively correlated to gestures.

In our method, to generate wide variety of gestures from the same text, but different context, we propose an attention based encoding 
technique, where the attention weights are estimated from the input word features embedded by BERT and multiplying them by each word feature.
The attention weight network is trained by using manually annotated ground truth information of TED 
Gesture-Type Dataset~\cite{TWT22}, which is expanded by us and make it publicly available after acceptance.
In addition, ACT2G can also generate arbitrary gestures by manually setting attention 
weight to specific word, where users want to emphasize by the gesture. 

The main contributions of our work are the following:
\begin{itemize}
\ifx
    \item Key-pose based gesture clustering with VAE for many-to-one mapping of text and gesture.
    
    \item attention-based text encoder to focus on words that should represent gestures.
    
    \item Contrastive learning for constructing the multi-modal space between gestures and text
\fi
    \item Contrastive learning for constructing the multi-modal space between 
	  gestures and text to achieve semantic gesture generation is proposed
    \item Attention-based text encoder to focus on specific words which represent 
	  gestures is proposed
    \item Attention-based gesture generating tool based on manual selection of 
	  keyword for content creators is developed.
    \item New gesture database including attention information is created and free for public use.
    %and  planned publicly available.
\end{itemize}

%

% Head 1
\section{Related Work}
Gestures can be divided into %considered to fall into 
mainly four categories, such as beat, deictic, iconic, and metaphoric~\cite{MCN94}. Beat is a gesture that has nothing to do with the content of the utterance; it is a gesture like shaking  arms in time with the inflection of the voice. The other three types of gestures are, respectively, pointing gestures, gestures for concrete objects or actions, and gestures for abstract concepts; these are called representational, which express the content of speech. Absaliev \textit{et al.} denote representational gestures as expressive gestures and analyze the connections between language and expressive gestures. \cite{abzaliev2022}. Deep Gesture Generation~\cite{TWT22} proposed a method for generating gestures that takes these gesture types into account, however this paper focuses on the generation of representational gestures.

With the remarkable development of deep learning, recent research in gesture generation has tended to be data-driven, with some methods generating gestures from audio, text, or both, or other modalities as well.
The trend in gesture generation in recent years has been probabilistic generative models, e.g., adversarial model~\cite{S2G19, FERSTL2020117, 3dconvgesture_2021},  normalizing flow-based model~\cite{AHK20}, and VAE model~\cite{A2G21}.
On the other hand, deterministic models like RNN-based models~\cite{TDN21, biLSTMA2G}, Seq2Seq model~\cite{YKJ19}, and auto-encoder model~\cite{Lu2021DoubleDCCCAEEO} also exist. 
Li \textit{et al.} \cite{A2G21} pointed out that deterministic generative models to date have been trained with a one-to-one mapping of audio or text to gesture, but because of the diversity of gestures, they trained to separate latent features so that text and gesture were one-to-many. Also, there is a model that predict gesture parameters from speech and refer to appropriate gestures from a database~\cite{EG21}. Outputting gestures directly from the database enable the generation of gestures that are more human-like. However generating representational gestures from audio alone is difficult because it is hard for the network to learn semantic information related to gestures. 

While there are many methods for generating gestures from audio\cite{S2G19, taras2019, A2G21, Ao2022, xu2022freeform, disco2022}, there are also methods for generating gestures from text, such as Seq2Seq~\cite{YKJ19}, transformer model~\cite{BRB21}, and GPT model~\cite{gao2023gesgpt}. A more recent trend is to generate gestures using both text and audio~\cite{GES20, Liang_2022_CVPR, Ao2023GestureDiffuCLIP}, using the speaker's ID as input~\cite{YBJ20, HA2G22}, and also facial expressions and emotions~\cite{liu2022beat}. Ginosar \textit{et al.} focused on generating individual-specific gestures because of the diversity of gestures~\cite{S2G19}, Yoon \textit{et al.} controlled the generated gestures by providing the speaker's ID as input~\cite{YBJ20}. For those who actually design the gesture, the more input modalities, the higher hurdle to generate gestures. Therefore, we propose a method to generate gestures using only text as input. Our method can also generate a gesture by adding Attention as an input, specifying the words that we want to appear as an representational gesture.

\section{Proposed Method}
ACT2G takes text as input, and output a realistic gesture. The training process is divided into three parts: (1) Gesture-VAE, (2) attention-based text encoder, and (3) contrastive learning. The ACT2G pipeline including (2) and (3) is shown in Fig. \ref{fig:network}. In Sec. \ref{sec:g_vae}, we introduce the gesture clustering as a preliminary preparation. Sec. \ref{sec:text_encoder} introduces the first half of ACT2G, attention-based text encoder, and Sec. \ref{sec:contrastive_learning} describes the contrastive learning process for gesture generation.

\begin{figure}[t]
    \centering
    \includegraphics[width=\linewidth]{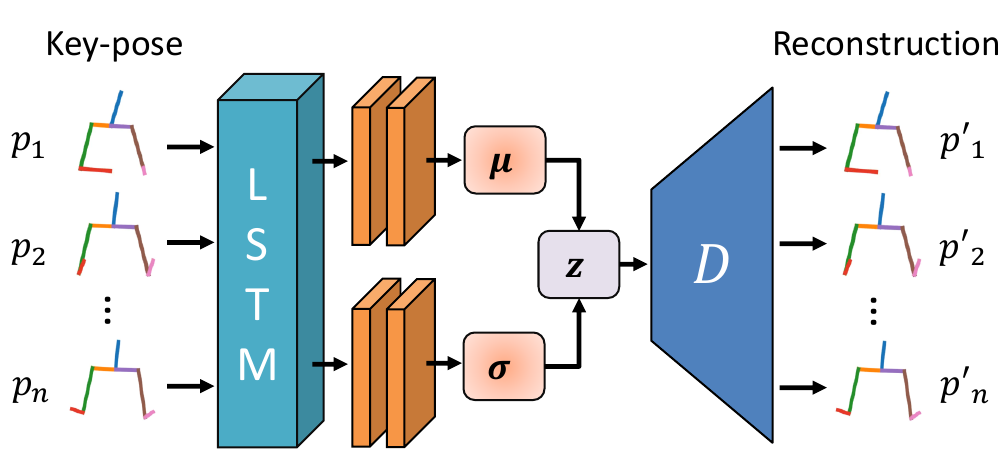}
    \caption{Gesture-VAE Network. The network takes the key-poses of the gesture as input and predicts the gesture feature $\textbf z$.}
    \label{fig:g_vae}
\end{figure}

\begin{figure*}[h]
    \centering
    \includegraphics[width=\linewidth]{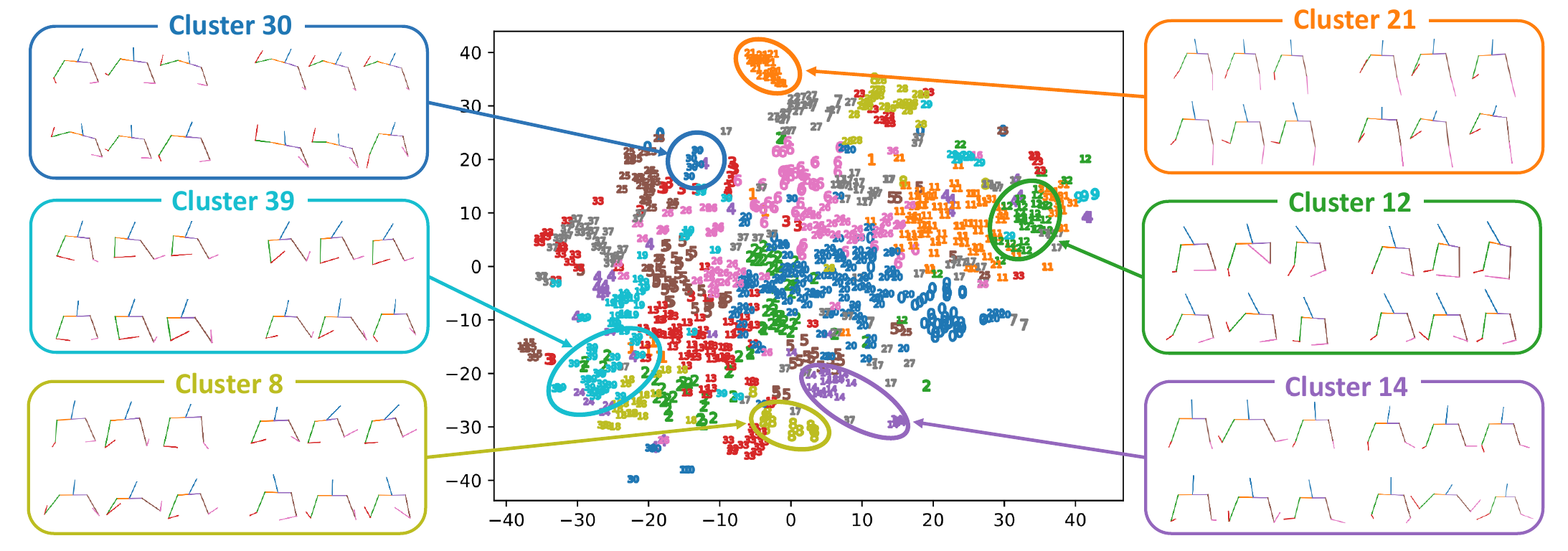}
    \caption{t-SNE visualization of gesture clustering results. The number of each data point represents the cluster number. Only clusters 30, 39, 8, 21, 12, and 14 show the 3 key-poses of four gestures.}
    \label{fig:g_vae_result}
\end{figure*}

%-------------------------------------------------------------------------
\subsection{Gesture Clustering Using Gesture-VAE}\label{sec:g_vae}
As Li \textit{et al.} mentioned~\cite{A2G21}, recent data driven methods take an approach where the input (text, audio, speaker ID, etc.) and the output (gesture) are mapped in a one-to-one fashion. However gestures are diverse, and the same gesture does not always appear in the same text. Inconsistencies in training data can hinder learning. Therefore, in order to achieve a one-to-many correspondences between gesture and text, we apply clustering on gestures in advance.  

We use intermediate features of VAE~\cite{Kingma2014} for the clustering of gestures. The network structure is shown in Fig. \ref{fig:g_vae}. One of the main problems when clustering time-series data such as gestures is the difference in sequence length. Especially in networks using RNN as encoders, problems such as vanishing gradient make feature extraction of too long data difficult. Therefore, we train the network using the key poses of the gestures as input. Considering key-poses such as labanotation plays an important role in the analysis of human motion, as shown in previous researches~\cite{Manoj2009KeyposeAS, IMY18}. We extracted the key-poses $\textbf p=\lbrace \textbf p_1,...,\textbf p_n \rbrace$ using off-the-shelf algorithm \cite{IMY18} in advance and input them into VAE. The key poses are input to the bi-directional LSTM, where the mean $\mu$ and standard deviation $\sigma$ are estimated and the latent feature $\textbf z \in \mathbb{R}^{32}$ is randomly sampled from the normal distribution corresponding to that parameter. 
The number of dimensions of $\textbf z$, 32, was determined empirically by plotting ellipses from the mean $\mu$ and standard deviation $\sigma$ and observing the overlap.
Then the gesture is reconstructed by decoder $D$ from $\textbf z$. The number of key poses $n$ ranged from 5 to 12 frames, and data were selected from the representational gestures in the TED Gesture-Type Dataset. Each key pose $\textbf p_t$ is represented as relative positions of the 8 upper body joints. The loss function for Gesture-VAE is as follows:

\begin{equation}
\mathcal{L}(\theta, \phi)=\mathbb{E}_{\mathbf{z} \sim \mathrm{q}_\theta(\mathbf{z} \mid \mathbf{p})}\left(\log \mathrm{p}\left(\mathbf{p}^{\prime} \mid \mathbf{z}\right)\right)-D_{K L}\left(\mathrm{q}_\theta(\mathbf{z} \mid \mathbf{p}) \| \mathrm{p}_\phi(\mathbf{z})\right)
\end{equation}

where $D_{KL}[.]$ denotes the Kulback-Leibler divergence, $\mathcal{L}$ refers to the likelihood of the parameters of encoder and decoder (i.e., $\theta$ and $\phi$) and $\textbf p^{\prime}$ denotes the result of reconstructing the key-poses. Latent feature $\textbf z$ are then used for gesture clustering. Gestures were clustered into 40 clusters by K-Means algorithm. 

The results of clustering the gestures are shown in Fig. \ref{fig:g_vae_result}. By extracting gesture features in VAE, each gesture was mapped and clustered in continuous space. For example, cluster 39 and cluster 12 are in opposite positions in Fig. \ref{fig:g_vae_result}, with the gesture of cluster 39 facing left, but the gesture of cluster 12 facing right. In addition, the gesture in cluster 21 has the left hand down, while the gestures in clusters 8 and 14 have the arm up gesture. 

Gesture and text in the same cluster were labeled Positive, while data in different clusters were labeled Negative. These positive and negative labels are used during contrastive learning in Sec. \ref{sec:contrastive_learning}.

%-------------------------------------------------------------------------
\subsection{Attention-based Text Encoder}\label{sec:text_encoder}
Designing gestures manually requires specialized knowledge and is very labor intensive. Gesture generation using AI simplifies this task. However these approaches cannot represent the gesture as the user would like to design it. Therefore, we propose a method to generate gestures by focusing on words that are likely to be expressed in gestures. Representational gestures in the TED Gesture-Type Dataset are annotated with the corresponding words in the utterance correspond to that gesture. We use this data to estimate attention weight $\textbf A$ that represents the weight of the word to focus on.

\begin{figure}[t]
    \centering
    \includegraphics[width=\linewidth]{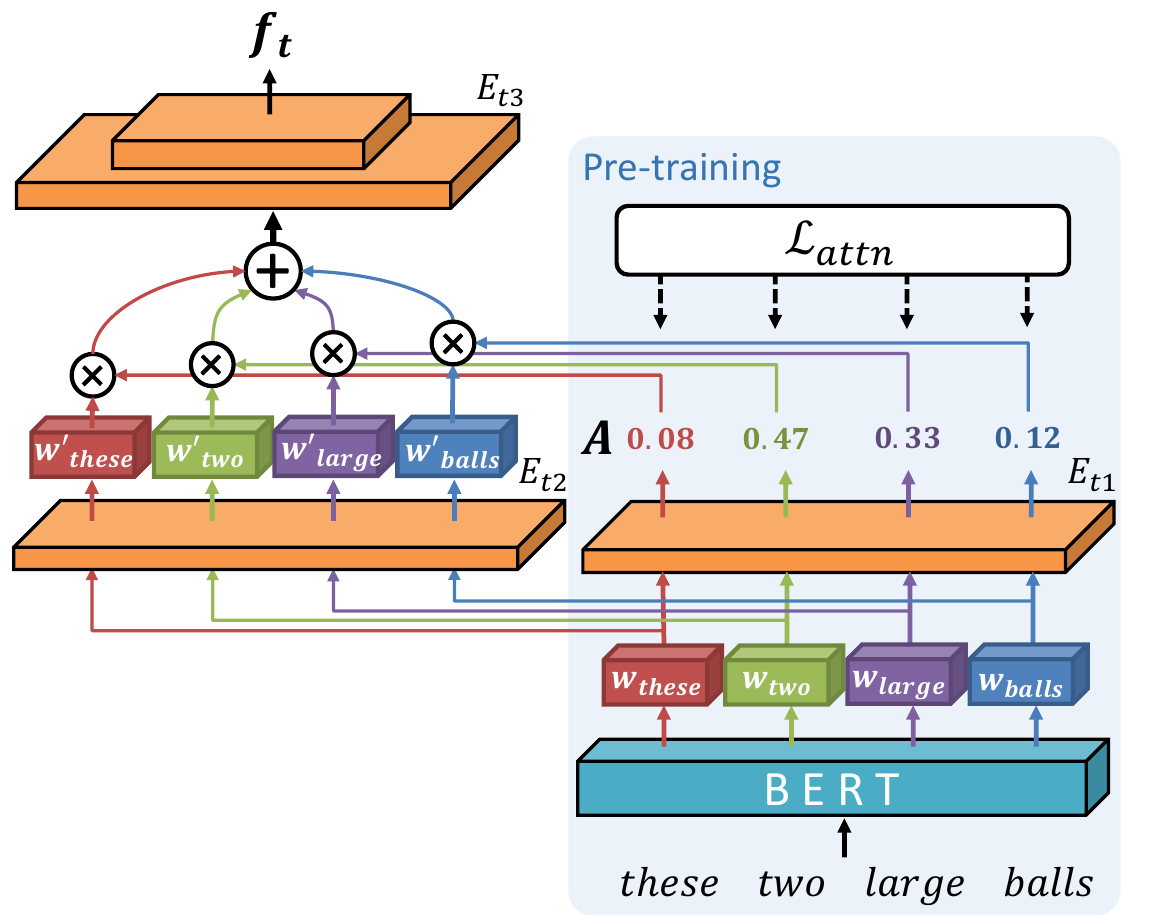}
    \caption{Attention-based text encoder. In this example, "these two large balls" is input and the words "two" and "large" are annotated as representing gestures.}
    \label{fig:attention_encoder}
\end{figure}

Our proposed network structure is shown in Fig. \ref{fig:attention_encoder}, which is the $E_t$ portion of Fig. \ref{fig:network}. First, each word is converted to a word embedding set $\textbf w=\lbrace \textbf w_1,...,\textbf w_T \mid \textbf w_i \in \mathbb{R}^{768} \rbrace$ by using pre-trained BERT~\cite{BERT}. $T$ is the number of input words, set to 32. Then, the attention weight $\textbf A=\lbrace A_1,...A_T \rbrace$ is estimated by encoder $E_{t1}$ from $\textbf w$, which consists of a fully connected layer: 

\begin{equation}
\mathbf{A} = f\left(E_{t1}\left(\mathbf{w}\right)\right),
\end{equation}

where $f$ is the following normalization function to ensure that text features are not affected by word count:

\begin{equation} \label{equ:normalize_func}
f_i(x)=\frac{x_i}{\sum_j x_j}.
\end{equation}

Another word feature $\textbf w^{\prime}$ is estimated from $\textbf w$ by using encoder $E_{t2}$ and by multiplying with $\textbf A$. 
During training, $\textbf w$ is always the same value for the same word because BERT is frozen, but $\textbf w^{\prime}$ is fine-tuned by $E_{t2}$.

Each word feature is then concatenated and the encoder $E_{t3}$ outputs a text feature $\textbf f_t$ as in the following equations:

\begin{equation}
\mathbf{w^{\prime}} = E_{t2}\left(\mathbf{w}\right)
\end{equation}
\begin{equation}
\mathbf{f_t} = E_{t3}\left(\|_{i=1}^T \left( \mathbf{A} \odot \mathbf{w^{\prime}} \right) \right),
\end{equation}

where $\|_{i=1}^T$ represents the vector concatenation from 1 to T and $\odot$ represents the Hadamard product. Attention weight $\textbf A$ is regularized by binary cross entropy loss:

\begin{equation}
\mathcal{L}_{attn}=-\frac{1}{T} \sum_{i=1}^{T} \hat{A}_i \cdot \log A_i+\left(1-\hat{A}_i\right) \cdot \log \left(1-A_i\right),
\end{equation}

where $\hat{A}_i$ is a ground truth with label 1 for the word corresponding to the gesture and 0 for the others. The part on the right side of Fig. \ref{fig:attention_encoder}, illustrate estimation process of $\textbf A$. This part was pre-trained with data from all representational gestures in the TED Gesture-Type Dataset. When training contrastive learning, which is discussed in Sec. \ref{sec:contrastive_learning}, $\textbf A$ is fine-tuned by using text features $\textbf f_t$ to reconstruct the gesture. During inference, gestures are generated from text features $\textbf f_t$, where $\textbf A$ can be explicitly given. In Sec. \ref{sec:generation}, we describe an experiment in which $\textbf A$ is also input to generate a gesture.

%-------------------------------------------------------------------------
\subsection{Contrastive Learning for Multimodal Space Construction}\label{sec:contrastive_learning}
Many recent gesture generation methods output sequences of poses directly from the network, but they are often overly slow or jerky. We assume that the slow movement problem is due to the generator's RNN and autoregressive model. Therefore, we propose a method to create a gesture library and search for the appropriate gesture from the library. ACT2G generates gestures using contrastive learning, which is often used to improve text embedding~\cite{Kiros2014UnifyingVE} or image and video retrieval~\cite{MLS18, Bain21}. 

The overview of the network structure is shown at Fig. \ref{fig:network}. Key-poses are extracted from the gesture and input to encoder $E_g$, which consists of a bi-directional LSTM or FCN. $E_g$ outputs the gesture features $\textbf f_g \in \mathbb{R}^{32}$, and the gesture is reconstructed through the decoder similar to Gesture-VAE, described in Sec. \ref{sec:g_vae}. Contrastive loss is as follows:

\begin{equation}
\mathcal{L}_{contrastive}=\frac{1}{B} \sum_{B}\left[\frac{1}{2}(\mathbf{P} \odot \mathbf{D})^{2}+\frac{1}{2} \max (0, \mathbf{m}-(\mathbf{1}-\mathbf{P}) \odot \mathbf{D})^{2}\right]
\end{equation}

where $B$ is the batch size, and $\textbf P \in \mathbb{R}^{B \times B}$ is the positive matrix between each data defined by the gesture clustering described in Sec. \ref{sec:g_vae}, and is a square matrix of 1 if each data is positive and 0 if negative. $\textbf D \in \mathbb{R}^{B \times B}$ is the L2 distance matrix between text feature $\textbf f_t$ and gesture feature $\textbf f_g$. And $\textbf m \in \mathbb{R}^{B \times B}$ is the margin, which is a square matrix with all elements $m$. We set $m = 20$ in practice. This contrastive loss means, in the multimodal space, the distance between gesture-text pairs, defined as positive in Sec. \ref{sec:g_vae}, is small, while the distance between pairs, defined as negative, is large. The loss to reconstruct the key-poses is:

\begin{equation}
\mathcal{L}_{reconst}=\frac{1}{B} \sum_{i}^B\left(p^{\prime}_i-p_i\right)^2.
\end{equation}

The loss function for the entire framework is as follows:

\begin{equation}
\mathcal{L}= \mathcal{L}_{attn} + \alpha \cdot \mathcal{L}_{reconst} + \beta \cdot \mathcal{L}_{contrastive} .
\end{equation}

\begin{figure}[t]
    \centering
    \includegraphics[width=\linewidth]{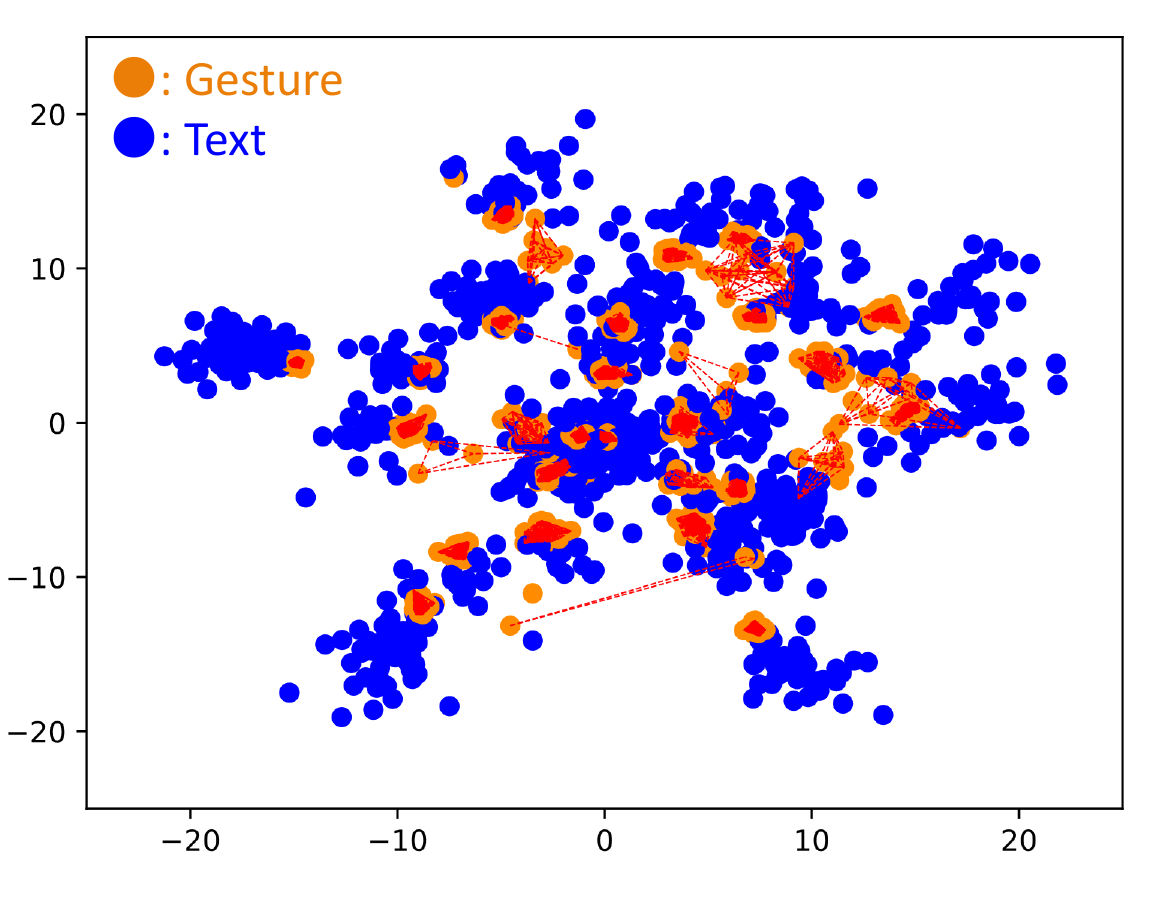}
    \caption{t-SNE visualization of multimodal space. The blue dots represent text features $\textbf f_t$, the orange dots represent gesture features $\textbf f_g$, and the red dash lines are mutually Positive gesture data pairs.}
    \label{fig:multimodal_space}
\end{figure}

Two parameters $\alpha$ and $\beta$ controls the weights of the loss terms, and they were empirically determined to 10 and 2, respectively. The multimodal space constructed by the contrastive loss is shown in Fig. \ref{fig:multimodal_space}. The blue dots, orange dots, and red lines represent text features $\textbf f_t$, gesture features $\textbf f_g$, and pairs of mutually Positive gestures, respectively. Even though all positive data are connected with each other by a red line, fewer of them are visible as a line compared to the number of dots, indicating that the data that are positive with each other are clustered together.

During inference, using reconstructed key-poses is difficult to humanize the gesture by simply interpolating between the key-poses. Therefore, we propose a method that uses the multimodal space to retrieve appropriate gestures from a gesture library. The gesture library contains the gestures in the training data and their corresponding positions in the multimodal space as shown in Fig. \ref{fig:multimodal_space}. When text is input, the text feature is extracted by the encoder introduced in Sec. \ref{sec:text_encoder}, and the gesture is randomly sampled from nearby that text feature in the multimodal space. Long input text is empirically divided into 8 word and entered into the network. Gesture speed is adjusted to match the length of the human voice, if present, or the synthesized voice, if absent. The gestures generated for each segmented text were combined by spline interpolation.

\section{Experiments}
In this section, we first introduce the dataset we used for training and evaluation in Sec. \ref{sec:dataset}.   
Then, in Sec. \ref{sec:ablation_study}, we describe a ablation study, and in Sec. \ref{sec:comparison}, we discuss the evaluation of gestures generated by our method and state-of-the-art methods.
Finally, a gesture generation tool with user specific attention mechanism is demonstrated in Sec. \ref{sec:generation}.

%-------------------------------------------------------------------------
\subsection{Dataset} \label{sec:dataset}
\paragraph*{Training ACT2G}
%Contrastive learning in ACT2G is meaningless without some correlation between gesture and text, because it finds commonalities between text and gestures and maps them into a multimodal space. 
The purpose of contrastive learning in ACT2G is to find correlation between texts and gestures and map them into a multimodal space. 
Therefore, it is necessary to train the network using representational gestures by excluding gestures that are unrelated to text, such as beat gestures. We, therefore used the TED Gesture-Type Dataset~\cite{TWT22}. TED Gesture-Type Dataset contains 13,714 gestures divided from TED videos, each annotated with three gesture types: beat, representational, or non-gesture. We used 4097 of these gestures, annotated as representational, for training. When pre-training the attention-based text encoder described in Sec. \ref{sec:text_encoder}, we used all 4097 gestures. 
And when training the entire ACT2G, we manually annotated most appropriate word representing the gesture for each text of 1000 gestures.
%And when training the entire ACT2G, we used 1000 gestures whose annotations we manually conducted. This is because some extra words were incorrectly annotated as corresponding gestures in the original data. 
For example, an representational gesture which has text of "something that made me very happy" and then pulling the arms back toward one's chest was annotated as "me."
%"made me very happy" in the original data. However, this gesture represents only "me", and "made" and "very happy" can be noise in contrastive learning. Therefore, 
Then, we used all the representational gestures for the attention-based text encoder pre-training, and used our original annotated data for the contrastive learning. 

\paragraph*{Evaluation}
Gesture evaluation methods, such as avatars to visualize gestures, question items, and the user interface used for evaluation, vary from method to method. In evaluating gestures, the user study, in which the evaluation is based on human perception, is the most important experiment to focus on, and it is important to evaluate the gestures in the same environment. 

Therefore, we used the widely used TED Gesture Dataset~\cite{YKJ19, HA2G22, TWT22} for the user study. We also used the Trinity Speech-Gesture Dataset~\cite{IVA18} for a verification of generalization, according to the GENEA Challenge~\cite{genea2020}.  
The TED Gesture Dataset is data from various people speaking, whereas the Trinity Speech-Gesture Dataset is data of a single speaker speaking on a variety of topics. 

The TED Gesture Dataset is used for 2D pose estimation from video by OpenPose~\cite{openpose}, followed by lifting to 3D~\cite{3dbaseline}, or 3D pose estimation by Expose~\cite{ExPose:2020}. On the other hand, Trinity Speech-Gesture Dataset uses marker-based motion capture to collect pose information. We divided each dataset into 5 to 15-second sequences, separated by sentence units. Of these, 30 sequences were used for evaluation and 2 sequences were used as attention checks.

%-------------------------------------------------------------------------
\subsection{Ablation Study} \label{sec:ablation_study}

\begin{figure*}[h]
    \centering
    \includegraphics[width=\linewidth]{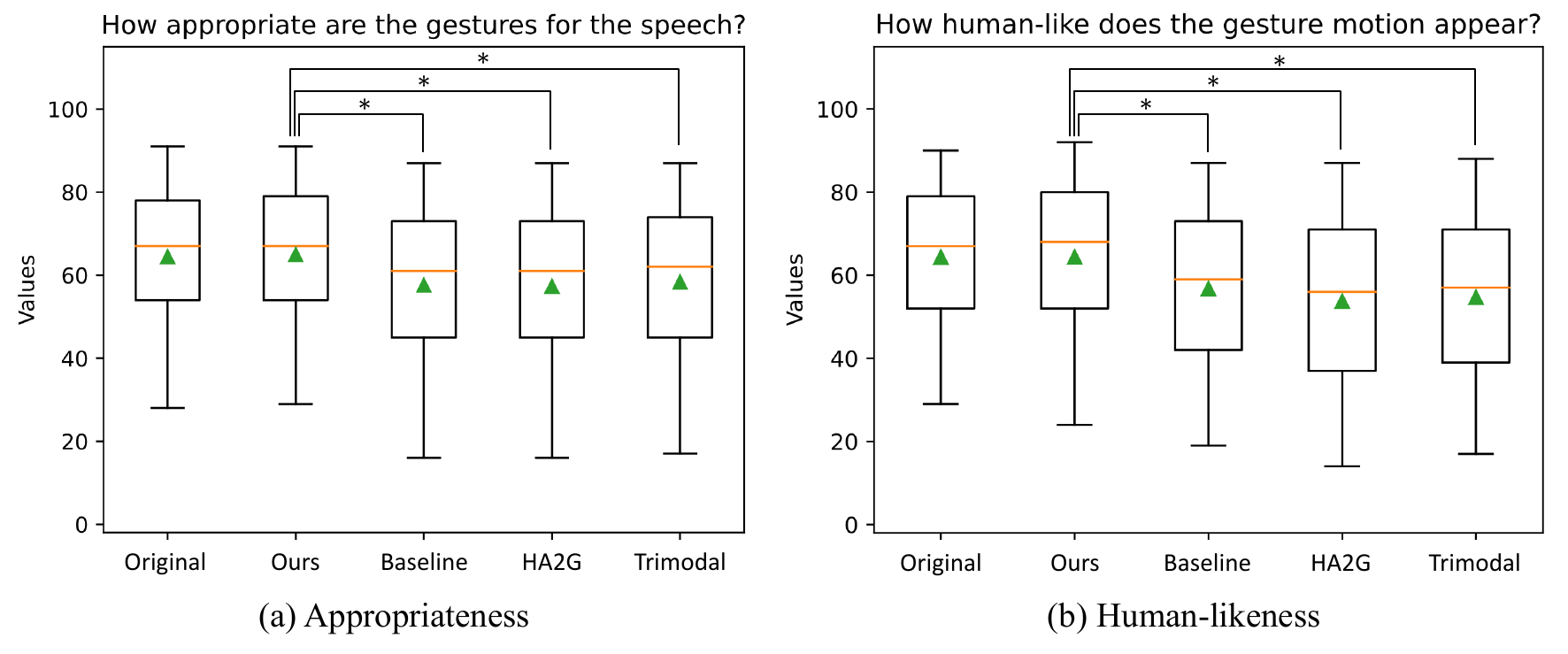}
    \caption{User study results with TED Gesture Dataset. ($*:p<0.01$) The orange line represents the median, and the green triangle represents the mean. Box edges are at 25 and 75 percentiles, while whiskers cover 95\% of all ratings for each.}
    \label{fig:user_study_result}
\end{figure*}

We first conducted an ablation study to gain more insights into our framework. 
We used Diversity score \cite{A2G21} and FGD score \cite{YBJ20} as metrics for quantitative evaluation and perceptual scores from the user study for qualitative evaluation. The Diversity proposed by Li \textit{et al.} is not suitable for computation with gesture data of different length, since the distance between gestures is computed with L1 distance of each joint in the same frame. We therefore used the distance between latent features trained in the Gesture-VAE introduced in Section \ref{sec:g_vae} as the distance between gestures:

$$
\text { Diversity }=\frac{1}{N \times\lceil N / 2\rceil} \sum_{a_1=1}^N \sum_{a_2=a_1+1}^N\left\|\mu_{a_1}-\mu_{a_2}\right\|_1,
$$

where $\mu$ refers to the latent vector of the VAE and $N$ is the number of motion clips.

The FGD proposed by Yoon \textit{et al.} \cite{YBJ20} could only handle short gestures of 34 frames. Therefore we extracted key poses from the gestures and used them as input to the Gesture-VAE (Fig. \ref{fig:g_vae}), allowing us to evaluate gestures with an average length of 233 frames and a maximum length of 1795 frames. Key poses were extracted using the method of Ikeuchi \textit{et al.} \cite{IMY18} method, which summarizes the entire motion. The gestures, the input to the VAE, were padded to 64 frames and the dimension of the latent space was set to 256 dimensions empirically. During the ablation study, we trained the VAE on the Trinity Gesture Dataset, and tested with 972 motion clips from TED Gesture Dataset.

In the user study, 50 participants in the Amazon Mechanical Turk rated 3 kind of gestures on a scale of 1-100 for the question "How appropriate are the gestures for the speech?". The requirements for participants in Amazon Mechanical Turk and the gestures used in the evaluation are the same as those described in Sec. \ref{sec:comparison} \textit{Comparison with Previous Methods} below.

Table \ref{table:ablation_study} shows the results of the ablation study. W/o contrastive is the case of no contrast learning, i.e., multimodal space is not created. After text features were extracted by BERT, gestures were selected by nearest neighbor method with the features from TED Gesture-Type Dataset. W/o attention means the way the gesture is generated without encoder $E_{t1}$ in Fig. \ref{fig:attention_encoder}. The result between w/o contrastive and full model in Diversity shows that it is possible to generate a variety of gestures by a multimodal space of text and gestures, rather than a simple nearest neighbor method with only text. The user study results show a statistically significant difference for the full model over the other two ways. The results with w/o contrastive were particularly significant, showing that the multimodal space created by contrastive learning allows for better gesture selection. Although the result of the full model was inferior to the other two methods in terms of the FGD, since FGD just means how similar the gesture is to the original gesture and it is not necessarily an appropriate gesture for speech, we prefer the user score and use the full model in subsequent evaluation.

\begin{table}[t]
\begin{tabular}{lllll}
                    & Diversity ↑       & FGD ↓             & User Score ↑ \\
\hline
w/o contrastive     & 34.99             & \textbf{155.91}   & $55.65 \pm 1.11 *$          \\
w/o attention       & 36.64             & 159.85            & $61.31 \pm 1.21 *$        \\
full model          & \textbf{37.43}    & 163.30            & $\mathbf{65.97 \pm 1.41}$  \\
\end{tabular}
\caption{Results of ablation study. $\pm$ means 95\% confidence interval, and $*$ means statistical differences from full model ($p<0.01$)}
\label{table:ablation_study}
\vspace{-0.5cm}
\end{table}

\begin{figure*}[t]
    \centering
    \includegraphics[width=\linewidth]{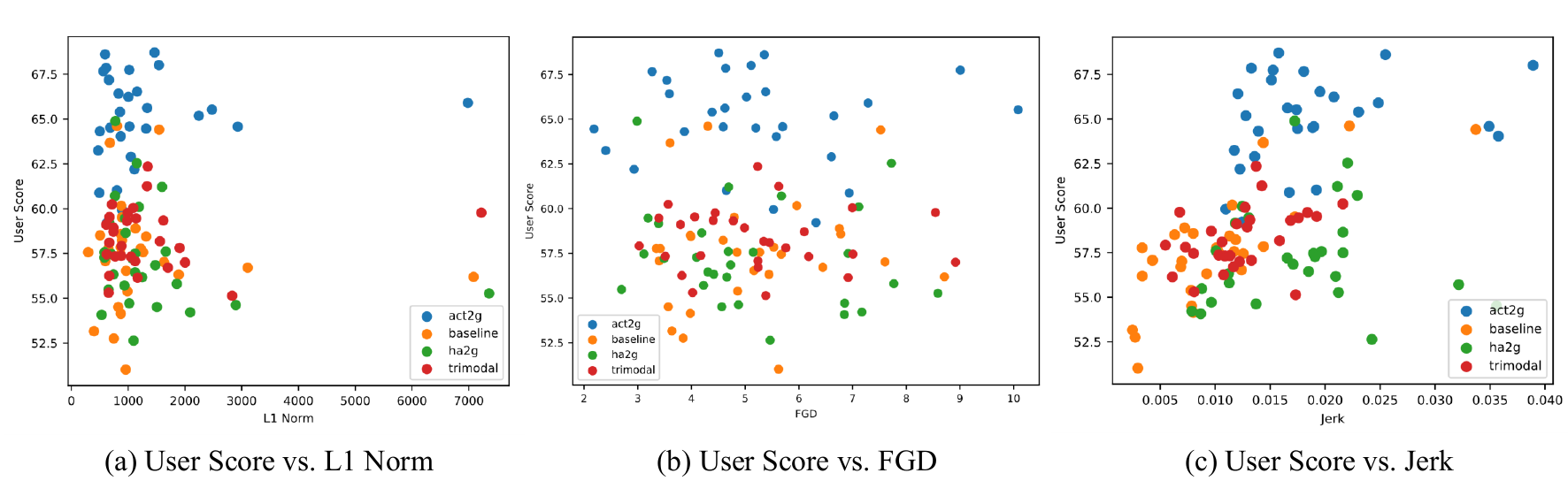}
    \caption{Relationship between user score and each metric. (a) User score vs. L1 norm. Correlation coefficient is -0.0305. (b) User score vs. FGD. Correlation coefficient is -0.0055. (c) User score vs. Jerk. Correlation coefficient is 0.4459.}
    \label{fig:score_analysis}
\end{figure*}

\subsection{Comparison with Previous Methods} \label{sec:comparison}
We conducted two user studies to compare gestures generated by ACT2G with existing methods and original gestures. 
The existing methods are compared with (1) Trimodal~\cite{YBJ20}, (2) HA2G~\cite{HA2G22}, (3) Deep Gesture Generation~\cite{TWT22} that serve as a baseline, since they are also considering semantic information for gesture generation. 
(1) \textbf{Trimodal} is a method that takes text, audio, and speaker identity as input and generates gestures using bidirectional GRU model. This model is trained with TED Gesture Dataset~\cite{YKJ19}. (2) \textbf{HA2G} is also a method that takes text, audio, and speaker identity as input and generates gestures using decoders that are hierarchically divided into body parts. When inferring, gestures are generated by only the audio and speaker identity as input without text. In both methods Trimodal and HA2G, speaker identities were chosen randomly from the training data, TED Gesture Dataset~\cite{YKJ19}. 
(3) \textbf{Baseline} takes text as input and uses a gesture library to generate gestures. This model generates gestures by each gesture-type generator after predicting the probability of gesture type (beat, representational, non-gesture) from each word. 
These methods were evaluated using test data in the TED Gesture Dataset.

% Evaluation envirionment (hemvip, AMT, questions, etc.)
We built an evaluation environment similar to the GENEA Challenge~\cite{genea2020}. As the user interface for evaluating gesture videos, we used HEMVIP~\cite{jonell_2021_hemvip}, which displays multiple videos in parallel and is intuitive and easy to use.  Also, the BVH Visualizer~\cite{genea2020} was used to visualize the gestures. Since gestures generated by ACT2G, Baseline and HA2G had to be converted to BVH format to use this Visualizer, each joint position was converted to Euler angle. The two questions we prepared for the gesture evaluation items are also the same as in GENEA Challenge, as follows:

\begin{itemize}
    \item[$(a)$] \textbf{Appropriateness} How appropriate are the gestures for the speech?
    \item[$(b)$] \textbf{Human-likeness} How human-like does the gesture motion appear?
\end{itemize}

When evaluating human-likeness, gestures were evaluated only on the basis of movement, with no audio. Study participants were recruited through the crowdsourcing platform Amazon Mechanical Turk instead of Prolific used in GENEA Challenge. The participants were selected if they satisfied the following three requirements: 1) they had completed more than 500 tasks, 2) their approval rate of task is over 90\%, and 3) they had passed the attention check. The attention checks are a way to check the quality of the worker, such as by displaying text such as "Attention! Please rate this video 35" in the some video or replacing the audio. 108 participants met the requirements in the Appropriateness and 115 in the Human-likeness. Participants rated each gesture on a 100-point scale labeled (from best to worst) "Excellent," "Good," "Fair" "Poor" and "Bad" in 20-point intervals. We randomly selected 28 gestures from those annotated as representational gestures in the TED Gesture-Type Dataset with an average original gesture's arm speed above a threshold.

\begin{figure*}[p]
    \centering
    \includegraphics[width=\linewidth]{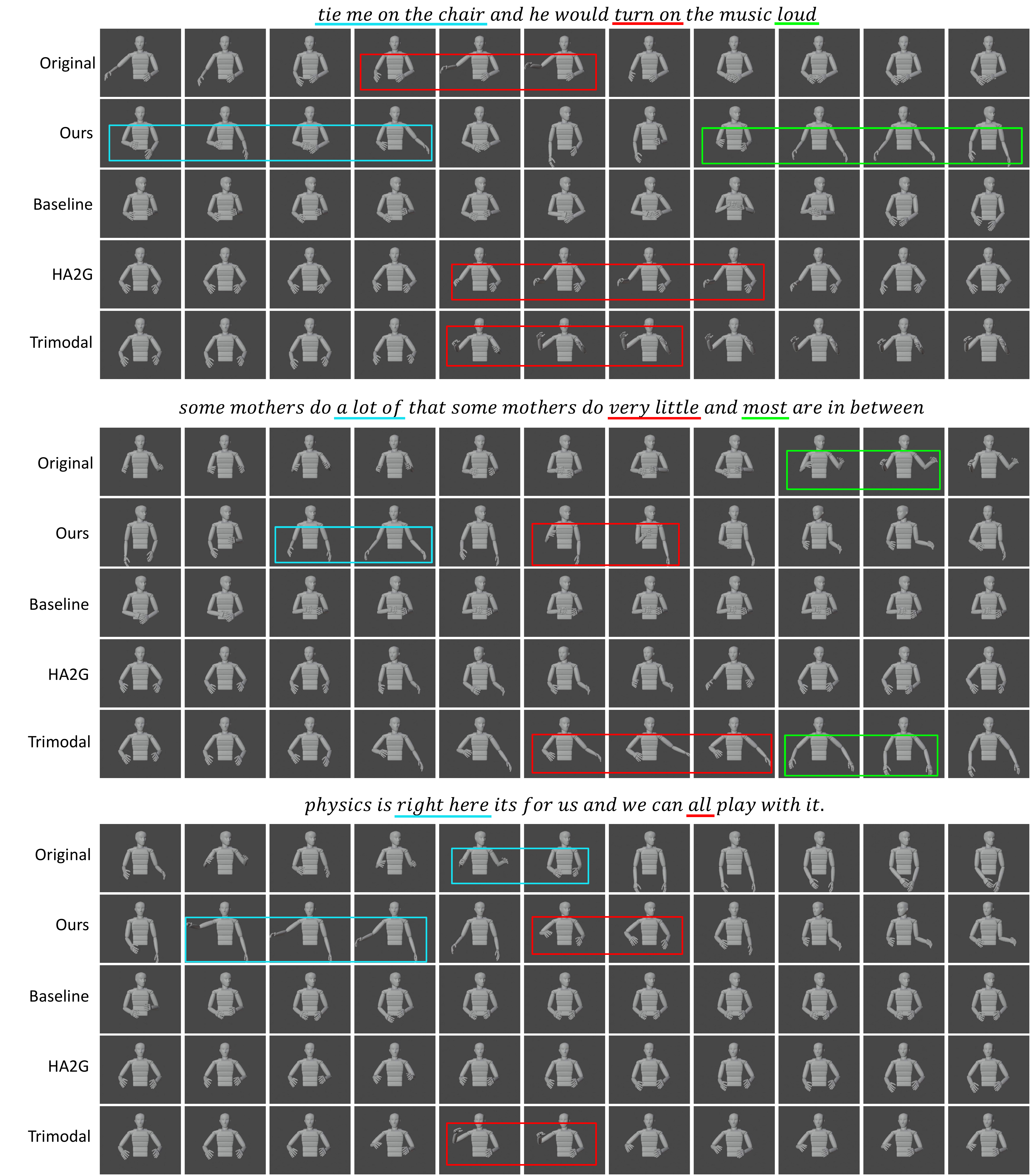}
    \caption{Qualitative results in TED Gesture Dataset. 3 sequences of input text and 5 kind of gesture pairs. Parts where the representational gestures seem to appear are marked with colored boxes and the corresponding text is also underlined in the same color.}
    \label{fig:generated_gestures}
\end{figure*}

Fig. \ref{fig:user_study_result} (a) shows the results of the "Appropriateness", Fig. \ref{fig:user_study_result} (b) shows the results of the "Human-likeness". The orange line represents the median value, and the green triangle represents the mean value. The results of ANOVA showed ACT2G was significantly higher than Baseline, HA2G and Trimodal with $p<0.01$ in both Apprppriateness and Human-likeness indicators. 
Examples of a generated gesture are shown in Fig. \ref{fig:generated_gestures}. In this figure, the parts where representational gestures might appear are indicated by colored texts and boxes. We can find that Ours has a higher frequency of representational gestures than the other methods. In the example in the upper row, original shows a gesture of spreading his right hand twice, which may represent "chair" or "turn on". On the other hand, our method makes a gesture corresponding to "tie me on the chair" by rotating the left hand, and a gesture corresponding to "loud" by spreading the arms wide. Baseline shows a movement like putting something down, while HA2G and Trimodal show a movement like spreading the right hand and raising both hands to shoulder level when saying "turn on," respectively. 
In the example in the middle row, original shows the movement of opening both hands when saying "most". Our gesture shows the movement of both arms spread wide when saying "a lot of" and the movement of the right hand to the center when saying "very little". Baseline generates a non-gesture, almost no hand movement. HA2G is gently tightening and opening the elbows, while Trimodal is spreading the left hand down at the "little" part and then opening both arms at the "most" part. 
In the example in the bottom row, the original gesture is a beat-like gesture with "here", "us", and "all" each with the arms swinging down. In ours, the right hand spreads when saying "right here" and the right elbow rises when saying "all". Baseline and HA2G are both very slow beat gestures. Trimodal is also slow motion and raises the right hand at the "all". The slow movement of previous methods is considered that GRUs and autoregressive models may have caused excessively smooth movements. As for HA2G and Trimodal, there may be some influence of randomly selected speaker identities. 

One major issue in gesture generation research is how to establish evaluation criteria. We focused on user scores to evaluate gestures according to the GENEA Challenge \cite{genea2020, genea2021, genea2022}, but it is very labor intensive to evaluate each generated gesture with a user study each time. Therefore, many previous studies have quantitatively evaluated gestures, and sought better evaluation metrics. In this study, we report on the better metric in this experiment by comparing how close the previously proposed metrics are to the distribution of user scores. We investigated three metrics used in previous studies: L1 norm \cite{S2G19, A2G21}, FGD \cite{YBJ20, Liang_2022_CVPR, HA2G22}, and Jerk \cite{taras2019, genea2020}. When calculating L1 norm, the speed of the generated gestures were normalized to make the duration as same as the original gestures. Fig. \ref{fig:score_analysis} shows the relationship between user score and each metric, where the user score means the result of the {\it Appropriateness} described in Sec.~\ref{sec:comparison}. %Comparison with Previous Methods. 
X-axes of \ref{fig:score_analysis}(a) and (b) are the L1 norm and the FGD, respectively, %in Sec.~\ref{sec:qualitative_evaluation}.
%, which are metrics of similarity to GT, 
and the smaller the better. 
The correlation coefficients between L1 norm, FGD and user score are -0.0305 and 0.0055, respectively, indicating little correlation. On the other hand, the correlation coefficient for jerk is 0.4459, indicating the certain correlation between them. 
%The reason for the trend toward higher ratings with larger jerks may be that the data used in this experiment consisted only of representational gestures, therefore there were many texts in which gestures with larger movements were expected to appear. Note that this is the result under the present experimental conditions and does not necessarily mean that L1 norm and FGD are irrelevant to gesture appropriateness.
We thought that the reason for the trend toward higher ratings with larger jerks might be explained by simple fact that meaningful texts are usually spoken by large movement of human.

\subsection{Verification of Generalization}\label{sec:generalization}
In Section \ref{sec:comparison} we evaluated four different models trained on the TED Gesture dataset on test data from the same dataset. While the TED Gesture Dataset contains data from a variety of speakers speaking on a variety of topics, the situation in which they are speaking in front of an audience is the same for all data. In addition, the test data did not include anything other than representational gestures such as beat gesture and non-gesture. Therefore, we conducted an additional user study to confirm the generalization performance.

For the test data, we used the Trinity Speech-Gesture Dataset~\cite{IVA18}, which is mo-cap data from a single speaker speaking on a variety of topics in the experimental room. The comparison methods include ours, baseline, and HA2G trained with the TED Gesture Dataset, plus Gesticulator \cite{GES20} trained with the Trinity Speech-Gesture Dataset. The experimental environment, including the interface for evaluation and the gesture visualization method, was the same as in the user study in Sec. \ref{sec:comparison}, and only human-likeness was used as a question item for participants. We randomly sampled 30 gestures from test data, and 115 participants rate the gestures.

\begin{figure}[t]
    \centering
    \includegraphics[width=\linewidth]{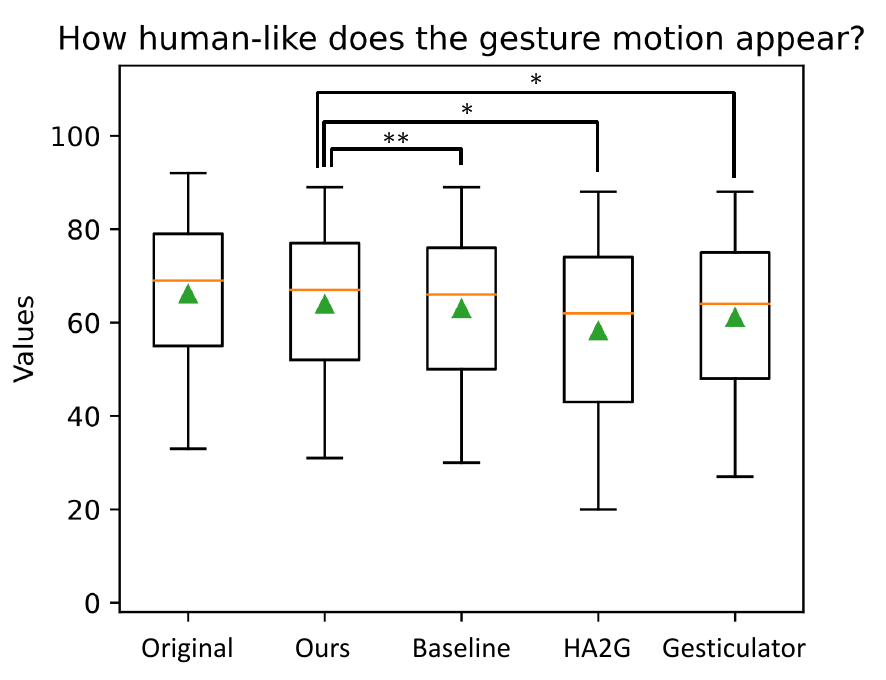}
    \caption{User study results with Trinity Gesture Dataset. ($*:p<0.01, **:p<0.05$) }
    \label{fig:generalization_user_study}
\end{figure}

Fig. \ref{fig:generalization_user_study} shows the results for the user study with Trinity Speech-Gesture Dataset. Statistical tests showed Original gesture was significantly best for any other gesture at $p<0.01$. Ours was significantly superior to Baseline at $p<0.05$, and to HA2G and Gesticulator at $p<0.01$. As can be seen from Fig. \ref{fig:score_analysis}, users tend to prefer the fast-moving (larger movement) gestures when they see five different gestures in parallel. Our method tends to generate fast-moving representational gestures, which is why it was rated higher than the baseline and other methods.
Gesticulator outperformed HA2G by p<0.01, possibly because HA2G is a model that is trained on the TED Gesture Dataset, while Gesticulator is a model trained on the Trinity Speech-Gesture Dataset.

%-------------------------------------------------------------------------
%\paragraph*{Attention-controlled gesture generation}
\subsection{Attention-controlled gesture generation}\label{sec:generation}
The attention-based text encoder described in Sec. \ref{sec:text_encoder} predicts attention, the likelihood that a gesture will appear, for each word. We assumed that by pre-defining the attention corresponding to each word and inputting it into the network, the gesture corresponding to the word with the highest attention weight would appear. Attention weights were set to 0.5 for specified words and 0.1 for other words.

\begin{figure*}[t]
    \centering
    \includegraphics[width=\linewidth]{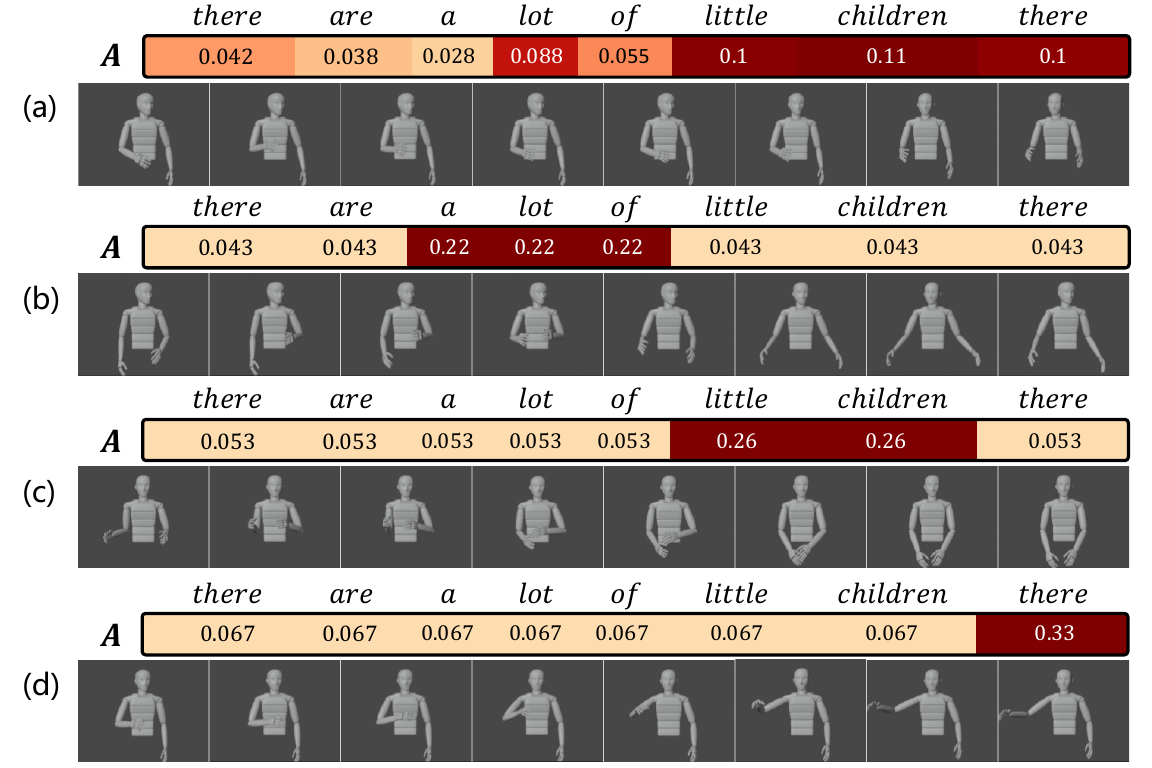}
    \caption{Qualitative results of gestures generated by inputting attention. (a) Case in which the network predicts $\textbf A$. (b) Case with high weight for "a lot of". (c) Case with high weight for "little children". (d) Case with high weight for the second "there".}
    \label{fig:attention_results}
\end{figure*}

Fig. \ref{fig:attention_results} shows the results with the text "there are a lot of little children there. Fig. \ref{fig:attention_results} (a) shows the result of attention weight $\textbf A$ estimated by the network with only text input, and (b), (c), and (d) are the results of generated gestures with higher weights for $\textbf A$ corresponding to "a lot of", "little children", and second "there", respectively. The each value of $\textbf A$ is normalized by Equation \ref{equ:normalize_func}, including the CLS token and padding tokens. In (a) gesture, the right hand is first rotated in a small motion, then both hands are brought forward. This appears to represent a "little" or a second "there". Also, (b), (c), and (d) gestures appear to represent "a lot of" with both arms outstretched, "little children" with arms down, and "there" with right arm extended, respectively. Because of random sampling from the gesture library, it may be generated slightly different gestures even for the same text and attention weight. Although a variety of gestures can be generated by entering text alone, even more diverse gestures can be generated by changing the attention weight.

\newpage
\section{Conclusions}
%Communication online and in virtual spaces has become more active, and the use of avatars has increased. Gesture generation is one challenge to make it more comfortable, however there was no method that takes into account which words in the text represent representational gestures to our best knowledge. 
Gesture generation recently becomes important research topic and many audio based techniques have been proposed, however, few on semantic information based.
In this paper, we proposed ACT2G, generating gestures from text, which includes three techniques: (1) gesture clustering based on latent space created by VAE;
%for one-to-many mapping of text to gesture based on gesture diversity, 
(2) an attention-based text encoder which explicitly considers words representing representational gestures, and (3) contrastive learning to retrieve content related gestures from the library. In the experiments, user studies were conducted confirming that our method outperformed existing methods in terms of "Appropriateness" and "Human-likeness." Another feature of the attention-based text encoder is that by manually setting the attention weight for each word, it is possible to generate gestures suitable for that word.

\paragraph{Limitations} 
ACT2G is primarily limited to three aspects. (i) This framework is trained on the TED Gesture-Type Dataset, therefore finger motion is not considered. (ii) Since ACT2G has only been trained on videos of a dozen seconds, a better interpolation method is needed to generate gestures for long sentence input. (iii) The TED Gesture-Type Dataset used for training only contains English videos, therefore gesture-type annotation is needed again for other languages.

% As a limitation, the framework is trained on the TED Gesture-Type Dataset, therefore, finger motion is not considered. 
%Attaching finger movements by training with dataset extended finger joints or synchronizing arm movements with audio is the future work.

% \vspace{-0.3cm}
\section*{Acknowledgment}
% \vspace{-0.1cm}
This work was supported by JSPS/KAKENHI JP20H00611, JP21H01457 and JP23H03439 in Japan.
% \vspace{-0.2cm}

\bibliographystyle{ACM-Reference-Format}
\bibliography{main}

% \printbibliography   

% %%
% %% If your work has an appendix, this is the place to put it.
% \appendix









\end{document}